\documentclass{aa}
\usepackage{graphicx,natbib}
\bibpunct[, ]{(}{)}{;}{a}{}{,}
\def\diff{{\rm d}}
\begin{document}
\title{BeppoSAX observation of the cluster Abell 970}


\author{G.B. Lima Neto\inst{1} \and H.V. Capelato\inst{2} \and
L. Sodr\'e Jr.\inst{1}  \and D. Proust\inst{3}
}

\offprints{G.B. Lima Neto,\\
\email{gastao@astro.iag.usp.br}}

\institute{%
Instituto de Astronomia, Geof\'{\i}sica e Ci\^encias Atmos. (IAG),
Universidade de S\~ao Paulo, Brazil 
\and
Divis\~ao de Astrof\'{\i}sica INPE/MCT, 12227-010, S\~ao Jos\'e dos 
Campos/S.P., Brazil
\and
Observatoire de Paris-Meudon, GEPI, F92195 Meudon, France}

\date{Received ????, 2002; accepted ?????, ????}

\abstract{ 
We report new results on the cluster of galaxies Abell 970 obtained from X-ray
observation with \textit{BeppoSAX}. Our analysis of the \textit{BeppoSAX} MECS
and LECS data in the range [0.15--10] keV reveals a mean cluster gas
temperature of $kT = 4.46_{-0.15}^{+0.14}$, a metallicity of $Z =
0.31_{-0.04}^{+0.05} Z_{\odot}$, and an interstellar hydrogen absorption
density column of $N_{\rm H} = 6.05_{-0.97}^{+1.29} \times 10^{20}$cm$^{-2}$. 
Moreover, we obtained azimuthally averaged radial profiles of these quantities.
Our results are consistent with the hypothesis that Abell 970 has been
disturbed by a past merger or by the ongoing merger process of a substructure,
that put the cluster out of equilibrium. This is also demonstrated by 
the offset between the gas and galaxy distributions. Combining the X-ray data
with a recently published analysis of new galaxy radial velocities, we
conclude that a subcluster 8 arcmin to the NW is falling into Abell 970
and will merge in a few Gigayears, thus disturbing Abell~970's newly acquired
equilibrium. The high $\alpha$-elements/iron ratio that we derive for this
cluster supports the hypothesis of early intracluster medium enrichment by
Type II supernovae.
\keywords{galaxies: clusters: Abell~970 -- clusters: X-rays -- clusters:
abundances} }

\maketitle
%

\section{Introduction}

Rich clusters of galaxies are the last structures to collapse in the Universe.
When isolated and within its virial radius, the cluster contents should
already be in quasi-stationary equilibrium by now. However, according to the
hierarchical structure formation scenario, rich clusters are formed by the
accretion of dark matter dominated haloes (\citealt{WhiteRees} and, e.g., the
$N$-body simulations of \citealt{Evrard}). These recurring merger events will
affect the cluster, moving it away from an equilibrium state. Therefore, the
study of the morphology and dynamics of cluster of galaxies can give valuable
information on structure formation in the Universe.

The hot ($kT \approx 2$--10~keV) and tenuous (central density, $n_{0}\approx
10^{-3} ${cm}$^{-3}$) X-ray emitting gas found in rich clusters of galaxies is
an excellent tool to probe the cluster properties and past history. As it was
already recognized almost three decades ago \citep{Lea}, the short relaxation
time of the gas, and the fact that the main emission mechanism is the
optically thin thermal bremsstrahlung, make it an excellent tracer of the
cluster gravitational potential well.

Besides the temperature and density, the X-ray observation of the intracluster
medium (ICM) can be used to measure the metal abundance (hereafter
metallicity) of mainly iron, but also of some $\alpha$-elements. Such
observations can contribute to our understanding of gas enrichment in
clusters, which depends on the early evolution of galaxies in clusters and
their interaction with the intra-cluster medium.

Here, we present an X-ray study of Abell 970 ($z = 0.0587$, richness class
$R=0$, B-M type III) following the work of \citet[Paper I hereafter]{Sodre}.

This cluster is in a rather dense environment, being part of the Sextans
supercluster (number 88 in the catalogue of \citealt{Einasto}, and number 378
in the catalogue of \citealt{Kalinkov}, which also includes Abell 978, 979,
and 993). It has moderate X-ray luminosity and it was observed with the {\it
Einstein} Imaging Proportional Counter \citep[e.g.][]{JonesForman}; it was
also detected in the \textit{ROSAT} Bright Survey Catalogue \citep{Schwope},
although a pointed observation has never been carried out with \textit{ROSAT}.
In the study of \citet{White97} using \textit{Einstein} IPC data, it has been
found that Abell 970 has at most a weak cooling flow, of about $20
M_\odot$~yr$^{-1}$ \cite[see also][]{Loken}.

In the dynamical analysis made in Paper I, we found that the galaxies in
Abell~970 have a mean velocity dispersion of 845~km~s$^{-1}$ (and up to $\sim
1000\,$km~s$^{-1}$ in the centre). This analysis also indicated that this
cluster must have substructure and is out of dynamical equilibrium. This
conclusion is also supported by an offset in the peaks of the surface density
distribution and the X-ray emission obtained from an \textit{Einstein} IPC
image. Further support came from a large-scale velocity gradient in the
cluster. We also found a discrepancy between the dynamical masses inferred
from the virial theorem and those inferred with the X-ray emission, which is
expected when the galaxies and the gas inside the cluster are not in virial
(or hydrostatic) equilibrium. Here we present new results on the distribution
of temperature, density, and abundances of the X-ray emitting gas, as well as
its $\alpha$-elements/iron ratio.

This paper is organized as follows. We describe in Sect.~\ref{sec:data} the
data obtained with \textit{BeppoSAX}. In Sect.~\ref{sec:results} we present
the X-ray fluxes and luminosities and determine and analyse the temperature
and metallicity radial profiles. The cluster gas and dynamical masses are
estimated in Sect.~\ref{sec:massa}. In Sect.~\ref{sec:merge} we discuss the
possible merging scenario of Abell 970, and in Sect.~\ref{sec:metal} we give
the individual metal abundances and discuss the consequences for the ICM
enrichment modelling.


\section{The Data}\label{sec:data}

\begin{figure*}[!htb]
 \centering
 \includegraphics[width=12cm]{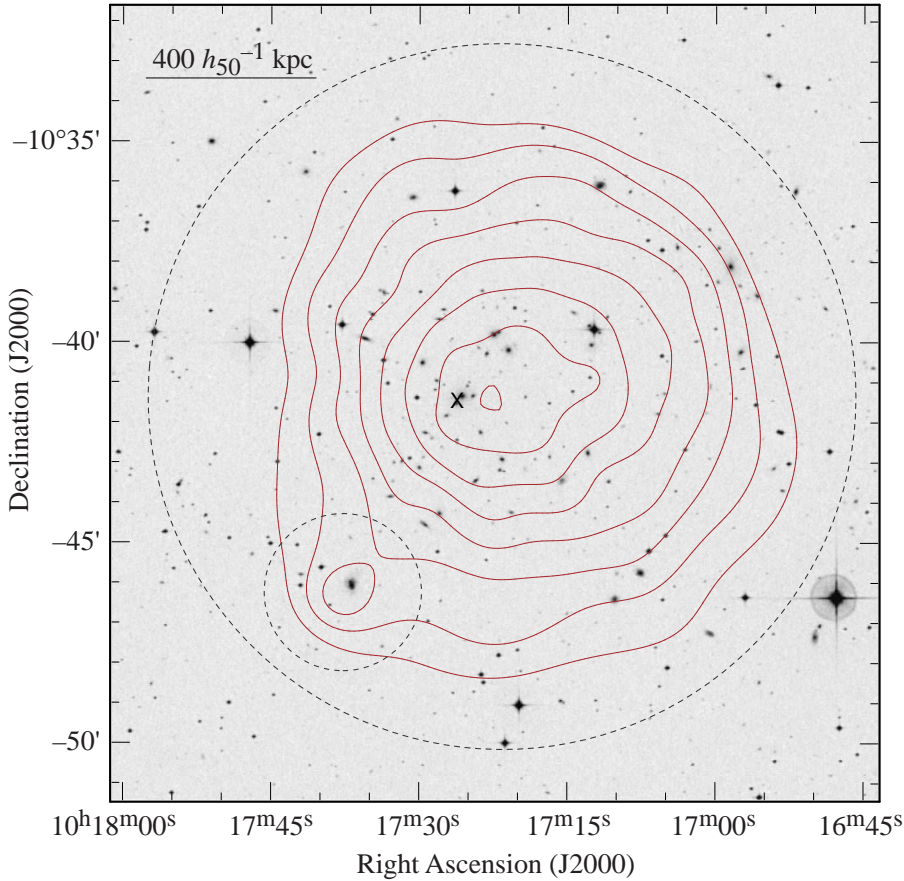}
 \caption[]{X-ray isocontours obtained with \textit{BeppoSAX} MECS 2 and 3
 superposed on a red DSS image (shown in grey scale). The larger dashed circle
 indicates the area used in our analysis; the smaller dashed circle indicates
 the excess X-ray emission tentatively associated with an active galactic
 nucleus (see text). The contours are spaced logarithmically. Notice the
 offset between the X-ray maximum and the centre of the galaxy distribution
 ($\alpha = 10^{\rm h}17^{\rm m}26.4^{\rm s}$, $\delta = -10^{\circ}41'27''$,
 J2000, marked with an ``\textsf{X}'', see also Fig.~\ref{fig:MECS23_a970.sec},
 below).}
 \label{fig:MECS23_a970_smooth}
\end{figure*}

Abell~970 was observed in July 2000 by the \textit{BeppoSAX} satellite
\citep{Boella1} with two of the narrow field instruments: the LECS providing
sensitivity in the [0.1--4.0]~keV band \citep{Parmar97}, and the MECS 2 and 3,
which are sensitive in the [1.6--10.5]~keV range \citep{Boella2}. The net
exposure times were $97\,289$ and $33\,619$~s for the MECS and LECS,
respectively.

We restricted the scientific analysis to the 8~arcmin ($\sim 760\,
h_{50}^{-1}\,$kpc at the distance of Abell 970) circle centered at the aim
point of the LECS and MECS instruments in order to prevent severe vignetting
effects \citep{Cusumano}. Moreover, the MECS entrance window is supported by a
beryllium ``strongback'' that is opaque to low energy photons ($E \la 5\,$keV)
and is not taken into account in the effective area files (Fiore, private
communication). The strongback structure affects detector positions between
$7.9'$ to $13.3'$ \citep{Cusumano}. Since we are analysing data inside the
MECS window structure, we opted for combining both MECS 2 and 3 event files in
order to improve the signal-to-noise ratio, instead of analysing them
separately as \cite{deGrandi02}, which would be essential for an analysis
extending beyond $\sim 8'$.

Figure \ref{fig:MECS23_a970_smooth} shows the X-ray isocontours
superposed on a ``Second Epoch Survey'' DSS optical image%
\footnote{The ``Second Epoch Survey'' of the southern sky was made by the
Anglo-Australian Observatory (AAO) with the UK Schmidt Telescope. Plates from
this survey have been digitized and compressed by the STScI UK Schmidt
Telescope and the Digitized Sky Survey.}.
The event file coordinates were corrected for the systematic error
reported by the \textit{BeppoSAX} team.
\footnote{cf.
\texttt{www.asdc.asi.it/bebbosax/coord\_correction.html}.}.
After correction, we verified a very good matching of both the MECS and LECS
images to the Einstein IPC image (see Paper I), including the small X-ray peak
at the SE, which, as pointed out in Paper I, is associated to a pair of
interacting cluster galaxies (numbers 37 and 38 of Paper I catalogue: $\alpha
= 10^{\rm h}17^{\rm m}37^{\rm s}$, $\delta = -10^{\circ}46'02''$, J2000). The
distance between the optical position of the pair and the associated X-ray
peak is less than $15''$.

A comparison of Figure~\ref{fig:MECS23_a970_smooth} with the \textit{Einstein}
IPC image (Figure 4 of Paper I) demonstrates that most of the features seen
here were already present in the IPC image. The peculiar offset between the
X-ray and the galaxy distribution maxima noticed before -- about 0.8 arcmin --
is also evident. As discussed in Paper I, this offset is a consequence of the
displacement of the X-ray isophotes in the direction of a concentration of
galaxies to the northwest ($\alpha \sim 10^{\rm h}16^{\rm m}48^{\rm s}$,
$\delta \sim -10^{\circ}38'$) which is centered on the brightest
cluster galaxy. This galaxy group however has no associated X-ray excess (see
also Fig~\ref{fig:MECS23_a970.sec}, below).

Given the moderate spatial resolution of the MECS and LECS detectors (FWHM 1.4
arcmin at 6~keV and 3.6 arcmin at 2~keV, respectively), we have defined
sub-regions in the field of view in order to perform separate spectral
analysis. The X-ray data in these regions were extracted with
\textsc{xselect}~2,
from the \textsc{HEAsoft} package%
\footnote{A  service of  the Laboratory  for High  Energy Astrophysics
(LHEA) at NASA/ GSFC and  the High Energy Astrophysics Division of the
Smithsonian Astrophysical Observatory (SAO).}.

We have defined 6 concentric annuli, each $1'28''$ wide (11 pixels), centered
at the X-ray maximum ($\alpha = 10^{\rm h}17^{\rm m}21^{\rm s}$, $\delta =
-10^{\circ}41'32''$). These regions are employed to derive the radial profiles
of various physical quantities. Notice that the last annulus extends over the
detector region affected by the MECS ``strongback''.

We have also defined 4 regions corresponding to the NW, NE, SW, and SE quadrants
(excluding the central region, see Sect.~\ref{sec:merge} for details). This
subdivision in sectors is very much like the one used by \cite{deGrandi}.
These regions, which are depicted in Fig.~\ref{fig:MECS23_a970.sec} below,
were chosen based on the galaxy distribution and their velocity distribution
(c.f. Paper I). Finally, we have defined a circular region of 2 arcmin (15
pixels) radius centered at the peak of the SE X-ray excess ($\alpha = 10^{\rm
h}17^{\rm m}37^{\rm s}$, $\delta = -10^{\circ}46'02''$, see
Fig.~\ref{fig:MECS23_a970_smooth}).

The X-ray spectrum of each region was modelled as being produced by a single
temperature plasma and we employed the \textsc{mekal} \citep{Kaastra,Liedahl}
model. For the whole cluster data, we also used the \textsc{vmekal} model
which allows the individual abundances for the elements to be fitted
independently. However, the abundance of some metals cannot be
constrained by the spectral fitting, therefore we have only used the following
elements: Si, S, Ar, Ca, Fe, and Ni. The others metals were fixed to $0.3
Z_{\odot}$.

\begin{figure}[htb]
 \centering
\includegraphics[width=8.6cm]{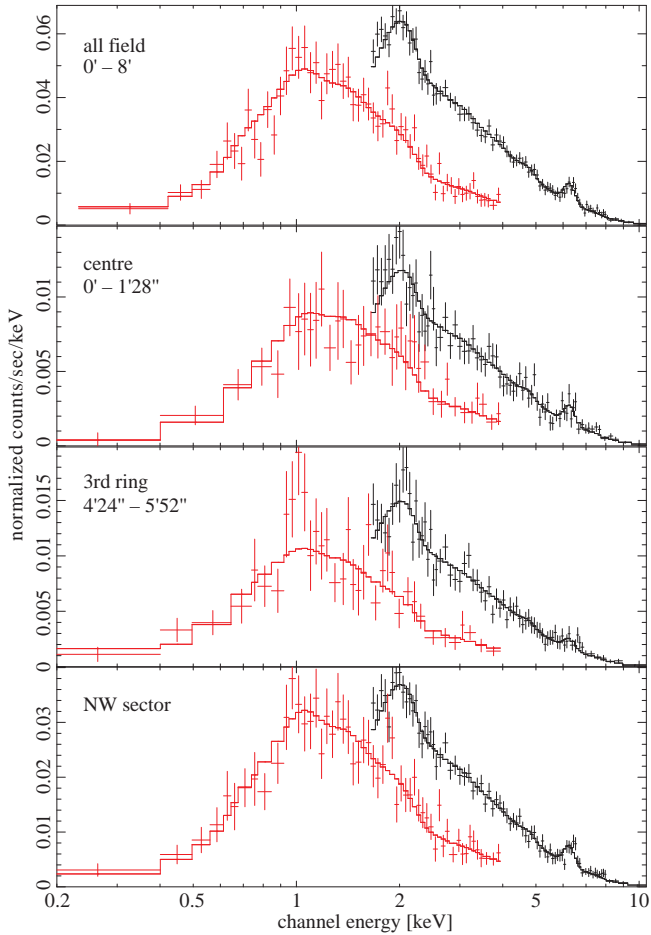}
 \caption[]{Four examples of the spectral fit. The left spectrum in each panel 
 is from the LECS, the right one from the combined MECS 2 and 3. The lines 
 passing through the data points correspond to the best fit absorbed MEKAL 
 model.}
\label{fig:spectra}
\end{figure}

The spectral fits were done using \textsc{xspec}~v11.0. Following the
\textit{BeppoSAX} ``Cookbook'' \citep{Fiore}, we have selected the data in the
energy range [0.12--4.0] keV for the LECS, and [1.65--10.5] keV for the MECS.
The photoelectric absorption -- mainly due to neutral hydrogen -- was computed
using the cross-sections given by \cite{Balucinska}, available in
\textsc{xspec}. Figure~\ref{fig:spectra} shows four examples of spectral fits.

For the last annulus, we have also done the spectral fits using only MECS data
in the interval [5.0--10.5]~keV, in order to check the masking effect of the
``strongback''. We verified that the difference between the best fit values
obtained either in the full energy range [1.65--10.5] keV or in the
[5.0--10.5]~keV range is well within the error bars (cf. below,
Sect.~\ref{sec:perfis}).

Background spectra were constructed from the event files (produced by the
\textit{BeppoSAX} Science Data Centre) of the standard background data, which
represent an assemblage of several empty fields devoid of any noticeable
source. For the MECS, we have used the combined background for detectors 2 and
3 available from the \textit{BeppoSAX} web site (under
\texttt{/pub/sax/cal/bgd/98\_11}). For the LECS, we have used the background
event file described by \cite{Parmar99} and available at
\texttt{/pub/sax/cal/bgd/99\_12}. The background spectra were extracted at the
same regions (in detector coordinates) defined above and used in the spectral
fits.

\section{Results}\label{sec:results}

\subsection{Flux and luminosity}

Table~\ref{tbl:Sumario} summarizes the fluxes and luminosities obtained for
the whole usable field and the central region. The bolometric X-ray luminosity
implies $T_{X} = 5.0 \pm 0.5$~keV based on the $L_{X}$--$T_{X}$ relation for
clusters obtained by \cite{Xue}.

\begin{table}[htb]
 \centering   
 \caption{Non-absorbed fluxes and luminosities in different energy bands. The
 energy band limits are given in keV.} \tabcolsep=0.75\tabcolsep
 \begin{tabular}{l  c c  c  c c}  
\hline
    Region & flux & flux &  $L_{X}$ & $L_{X}$ & $L_{X}$\\
       & [0.1--2.4] & [0.5--4.5] & [0.1--2.4] & [0.5--4.5] & bolom.\\ 
\hline 
   All field  & 27.8 & 30.0 & 42.2 & 46.3  & 83.1 \\
       Centre & 5.13 & 5.69 & 7.78 & 8.75 & 16.3 \\ 
\hline 
 \end{tabular} 
\label{tbl:Sumario}
\begin{flushleft}
   \vspace{-1ex}         
   Note: flux units are $10^{-12}$erg~cm$^{-2}\,$s$^{-1}$.\\ 
   Luminosity units are $10^{43} \, h^{-2}_{50} \,$erg~s$^{-1}$.
\end{flushleft}
\end{table}

As suggested in Paper I, the SE X-ray emission excess visible in
Fig.~\ref{fig:MECS23_a970_smooth} may be due to an active nucleus excited by
the merger of the pair of interacting galaxies. In order to verify this
hypothesis, we have extracted a spectrum in a 2~arcmin circle comprising this
feature and tested different spectral models (see Table~\ref{tbl:seyf}).

\begin{table}[htb]
 \tabcolsep=0.7\tabcolsep
 \centering
 \caption[]{Spectral fits of the south-east X-ray excess. $N_{H}$ is the 
 hydrogen column density in units of $10^{20}$cm$^{-3}$, $kT$ is the 
 temperature in keV, $Z$ is the metallicity, $\alpha$ is the spectral index 
 and ``dof'' is the degrees of freedom. Errors are at the $1 \sigma$ 
 confidence level.}
 \begin{tabular}{l c c c c c}
  \hline
  Model     & $N_{H}$ & $kT$ &  $Z/Z_{\odot}$ & $\alpha$ & $\chi^{2}/$dof \\
  \hline
  \textsc{pow} & $< 4.1$ & --- & --- & $1.5 \pm 0.1$ & 80.7/89 \\
  \textsc{mekal}  & $< 2.5$           & $15.2 \pm 7.5$   & $0.35^{\ddagger}$ & --- & 80.0/89 \\
  \textsc{mekal}  & $6.05^{\ddagger}$ & $14.2\pm 6.6$    & $0.35^{\ddagger}$ & --- & 81.1/90 \\
  \textsc{mekal}  & $< 50$            & $4.5^{\ddagger}$ & $0.55 \pm 0.29$   & --- & 94.1/89 \\
  \textsc{mekal}  & $6.05^{\ddagger}$ & $4.5^{\ddagger}$ & $0.58 \pm 0.33$   & --- & 95.3/90 \\
  \hline
 \end{tabular}
\begin{flushleft}
  \vspace{-1ex}
  ${}^{\ddagger}$ Value fixed.
\end{flushleft}
 \label{tbl:seyf}
\end{table}

We have tested a power-law (\textsc{pow}) and a thermal plasma
(\textsc{mekal}) models. A composite model, power-law plus \textit{mekal}
(i.e., a non-thermal source superposed on the cluster thermal emission) could
not be fitted unambiguously -- fits with different initial parameters ended
with different minimum $\chi^{2}$ in parameter space. Therefore, we
considered two opposite scenarios: I) the X-ray excess comes from a
non-thermal source and the emission inside a circle of 2 arcmin will be
dominated by a power-law spectrum; II) the merging galaxies have no AGN and
the emission is mostly of thermal origin.

We found that the best models, both statistically equivalent, are the
power-law and the thermal plasma endowed with the unreasonable high
temperature of 14~keV or more. Fixing the plasma temperature at 4.5~keV (i.e.,
about the cluster mean temperature, see below), whether or not keeping fixed
metallicities and $N_H$, produces poorer fits. We are thus led to suggest that
the X-ray source is of non-thermal origin, probably due to the nuclear
activity of one or both of the components of the interacting pair of galaxies
optically associated to it. This conclusion should however be regarded with
caution given the poor quality of the spectrum in this region, with net counts
rates of $5.1 \times 10^{-3}\,$count/s for the MECS and $2.3 \times
10^{-3}$count/s for the LECS.

\subsection{Radial profiles}\label{sec:perfis}

We have computed the radial profiles for several astrophysical quantities
using the circular annuli defined in Section~\ref{sec:data}, with the combined
MECS and LECS data. Since the SE X-ray excess may be of non-thermal origin, we
have done the spectral fits excluding and including this region. The results
are virtually the same, because the contribution of the SE excess to the X-ray
counts inside the annuli is not significant. The X-ray SE excess was
therefore excluded from the data prior to the spectral fits presented here.

\begin{figure}[htb]
 \centering
\includegraphics[width=8.6cm]{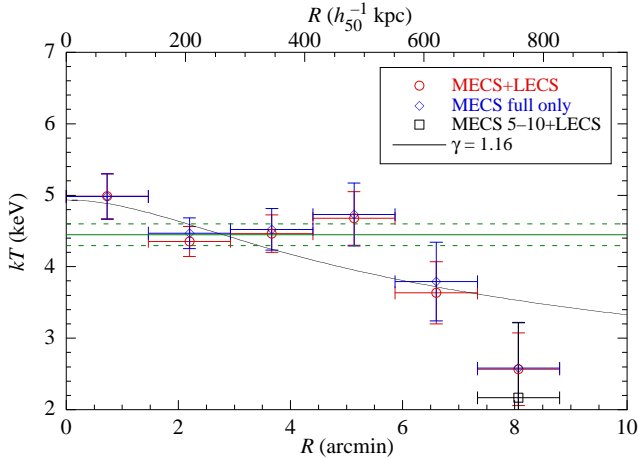}
 \caption[]{Temperature profile using either both LECS and MECS or only MECS
 data (which is more appropriate for cluster temperatures $T \ga 3\,$keV). For
 the last annulus there is an extra data point (square) where the spectral fit
 was done using only the MECS [5.0--10.5]~keV band. The error bars are at the
 $1 \sigma$ confidence level. The horizontal lines show the mean temperature
 (full line) and the $1 \sigma$ error bar (dashed) measured using all data
 within 8 arcmin ($kT = 4.46_{-0.15}^{+0.14}$). The curve (full line) shows
 the best fit of a polytropic model (see text).}
\label{fig:kT-R_A970Beppo}
\end{figure}

Figure \ref{fig:kT-R_A970Beppo} shows the resulting  
temperature profile
\footnote{Actually, it is the \textit{emission weighted temperature} that is
obtained in a spectral fit. However, for physical realistic gas models, the
difference between the actual gas (electron) temperature and the measured
emission weighted temperature should be negligible and well within the
observational uncertainties \citep{Komatsu}.}.

The temperature profile was also determined using only MECS data and fixing
$N_{\rm H} = 5.42 \times 10^{20}$cm$^{-2}$ (the galactic value,
\citep{Dickey}. This was done for two reasons: first, to check the robustness
of the fits, whether or not there was any spurious effect introduced by the
use of two different detectors.
 
The second reason is that when both the temperature and hydrogen column
density are taken as free parameters, and depending on the spectral range
being adjusted, the spectral fitting procedure may produce spurious
correlations between best-fit values of $kT$ and $N_{\rm H}$ [see, e.g.,
\cite{Pislar}].
 
As it can be seen from Fig.~\ref{fig:kT-R_A970Beppo}, the resulting
temperature profiles, that is, for MECS data only and $N_{\rm H}$ fixed at the
galactic value and for the combined MECS and LECS data and $N_{\rm H}$ free,
are almost indistinguishable. This demonstrates the internal consistency of
the modelled temperatures and a spurious correlation between fitted $kT$ and
$N_{\rm H}$ values is not present.

An additional fit for the outermost annulus was done using only the
[5.0--10.5]~keV MECS data (but still using the full LECS). This was done to
verify whether or not the MECS ``strongback'' has an important masking effect
for low energy photons. It turns out that the effect on the temperature
determination is within the error bars (which increases compared to the fit
using the full MECS range).

A simple parametric form of the ICM temperature can be obtained using the
polytropic equation of state. A polytropic model for the ICM is motivated by
the results of hydrodynamical simulations \citep[e.g.,][]{Suto}, theoretical
models \citep[e.g.,][]{Cavaliere,DosSantos}, and observed temperature
gradients \citep[e.g.,][, but see also \citealt{Irwin1,Irwin2,deGrandi02}
for a different observational result -- see discussion
below]{Markevitch,Finoguenov1}.

The polytropic gas is defined by:
\begin{equation}
    P \propto \rho^{\gamma} \, ,
    \label{eq:polytrop}
\end{equation}
where $P$ and $\rho$ are the pressure and density of the gas, respectively,
and $\gamma$ is the gas polytropic index. For an ideal gas, $P= n\, kT$ ($n$
is the number density of particles), therefore the temperature and density are
related by:
\begin{equation}
    kT \propto n^{\gamma - 1} \, .
    \label{eq:kTpolytrop}
\end{equation}

For the gas density profile, we use the well known $\beta$-model
\citep{Cavaliere1}:  
\begin{equation}
    n(r) = n_{0} \, \left[ 1 + 
    \left(\frac{r}{r_{c}}\right)^{2}\right]^{-3\beta/2} \, ,
    \label{eq:betamodel}
\end{equation}
where $n_{0}$, $r_{c}$ and $\beta$ are the central density, the core radius,
and the shape parameter. We adopt the values determined in Paper I: $\beta =
0.66\pm 0.44$ and $r_{c} = (260 \pm 20) h^{-1}_{50}$kpc or $2.8 \pm
0.2$~arcmin. For our considerations we have used the \textit{Einstein} IPC
results because this instrument provides a higher spatial resolution (about 25
arcsec) compared to the \textit{BeppoSAX} MECS (about 1.5 arcmin). The
temperature profile is then written as:
\begin{equation}
    T(r) = T_{0} \left[ 1 + \left(\frac{r}{r_{c}}\right)^{2} 
    \right]^{-3\beta (\gamma-1)/2} \, ,
    \label{eq:TempProfile}
\end{equation}
where $T_{0}$ is the central temperature. 

A least-square fit of Eq.~(\ref{eq:TempProfile}) yields $\gamma = 1.15 \pm
0.05$ and $T_{0} = 4.93 \pm 0.23$ with $\chi^{2} = 9.3$ for 4 degrees of
freedom. This value agrees well with those predicted by \cite{Komatsu} and
found by \cite{Finoguenov2}. The fit, shown in Fig.~\ref{fig:kT-R_A970Beppo},
is poor because of the very steep temperature gradient beyond 5 arcmin. In
their study of a sample of 21 nearby clusters observed with \textit{BeppoSAX},
\cite{deGrandi02} find that the temperature profiles of clusters are
characterized by isothermal cores extending to about one-fifth of the cluster
virial radius, followed by steep temperature gradients which are steepest for
the non-cooling flow clusters. These authors also find poor fits when trying
to adjust a polytropic law to the overall temperature profiles. This situation
is much similar to that observed in Abell~970 which, when compared to their
sample of non-cooling flow clusters, presents a steeper gradient. In
section~\ref{sec:mass} we use both the isothermal and the polytropic model to
estimate the total cluster mass.

We may check whether the last three radial points still correspond to a stable
state or whether the temperature gradient is too steep for stability. If we
take into account only the three external points, i.e., $5 < R < 8$ arcmin,
the polytropic index increases to $\gamma = 1.7 \pm 0.2$, which is slightly
above the ideal gas value, 5/3, suggesting that the outer region may be barely
in stable equilibrium.

\begin{figure}[htb]
    \centering
    \includegraphics[width=8.6cm]{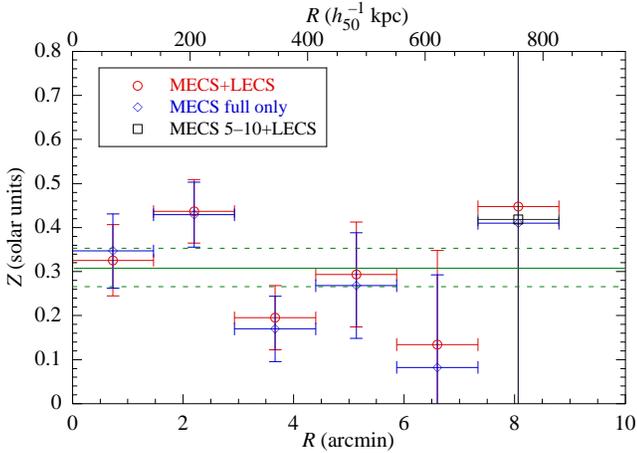}
    \caption[]{Metallicity profile using either both LECS and MECS or only
    MECS data (the metallicity is determined mainly by the iron complex at
    $\sim 6.8\,$keV). The square symbol has the same meaning as in the
    previous figure. The error bars are $1 \sigma$ confidence level. The
    horizontal lines show the mean metallicity measured using all data inside
    8 arcmin ($Z = 0.31_{-0.04}^{+0.05} Z_{\odot}$).}
    \label{fig:Z-R_A970Beppo}
\end{figure}

The \textsc{mekal} plasma model spectral fit also gives the ICM metallicity.
Figure \ref{fig:Z-R_A970Beppo} shows the metal abundance radial profile. In
order to check the fit robustness, we also did the profile determination using
only the MECS data, besides the fits using both detectors. Both profiles are
very similar, well within the error bars.

As before, we made an additional fit for the last annulus using only photons
with $E \ge 5\,$keV. Although the metallicity value obtained is close to the
spectral fit with the full MECS range, the error bars are so big that, in
fact, the metallicity should be considered undetermined at this last annulus.

The metallicity profile shown in Fig.~\ref{fig:Z-R_A970Beppo} has no apparent
gradient, being consistent with the mean value found for the whole cluster
(which is a typical value for a cluster of such richness and temperature).
Contrary to some other clusters, e.g., Abell~496 \citep[using ASCA
data]{Dupke} or Abell~85 \citep[using \textit{BeppoSAX} data]{LimaNeto}, this
cluster does not show an increase of metal abundance in the centre (defined by
the X-ray peak emission). \cite{Irwin3}, also using \textit{BeppoSAX} data,
find that clusters generally show a negative metallicity gradient, but this
gradient is usually smaller or even disappears in non-cooling flow clusters.

\begin{figure}[htb]
    \centering
    \includegraphics[width=8.6cm]{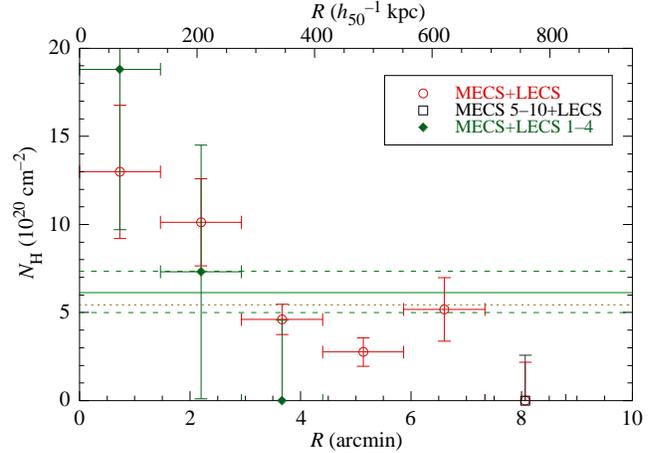}
    \caption[]{Hydrogen column density profile using LECS and MECS data. The
    error bars are $1 \sigma$ confidence level. Filled diamonds correspond to
    fits with LECS counts restricted to the range [1--4]~keV (but all MECS).
    The open square corresponds to the last annulus fit in which MECS counts
    were restricted to the range [5--10.5]~keV (see text). The horizontal full
    line shows the mean hydrogen column density measured using all data within
    8 arcmin; the horizontal dashed lines correspond to $1 \sigma$ error bars
    ($N_{\rm H} = 6.05_{-0.97}^{+1.29} \times 10^{20}$cm$^{-2}$). The
    horizontal dotted line indicates the mean galactic value, $N_{\rm H} =
    5.42 \times 10^{20}$cm$^{-2}$.}
    \label{fig:nH-R_A970Beppo}
\end{figure}

Figure \ref{fig:nH-R_A970Beppo} shows the radial profile of hydrogen column
density. For comparison we also show the galactic value at the position of
Abell 970. Here we observe a gradient, with significantly higher hydrogen
column density in the inner 2 arcmin. This excess is significant at a
2-$\sigma$ level, ignoring the uncertainty of the galactic $N_{\rm H}$ measure.

This level of significance may indicate that we are detecting an X-ray
absorption excess compared with the expected absorption predicted by the
observed galactic HI column density. The origin of such an excess is still
controversial \citep[and references therein]{Allen00}, the most promising
hypotheses being very cool molecular gas and/or dust grains. Moreover, the 
amount of this cool material depends on the details of the spectral modelling 
\citep{McCarthy02}.

Figure~\ref{fig:nH-R_A970Beppo} shows that $N_{\rm H}$ decreases outwards,
reaching the galactic value. However, a caveat must be introduced here: (I)
there may be variations of the galactic $N_{\rm H}$ that are unresolved by the
galactic survey at the cluster scale; (II) the $N_{\rm H}$ determination
relies almost exclusively on the LECS data (photon energy $\la 1.5\,$ keV).
The error bars in this case may be underestimated given the low count rate,
specially at the last bin.

Moreover, there are a number of issues associated with lowest energy photons
detected by the LECS: the spatial resolution is poor ($\approx
5.14/\sqrt{E_{\rm keV}}\,$arcmin) and there is a disagreement between the
measured (on-ground) and theoretical effective area for $E \la 0.3\,$keV
\citep[see][]{Parmar97}. Selecting only LECS counts with $E \ge 1.0\,$keV (the
spatial resolution is $\sim 5\,$arcmin at $E = 1\,$keV) can minimize these
effects. Figure~\ref{fig:nH-R_A970Beppo} shows the additional $N_{\rm H}$
determination with such a selection, for the centre and the first two rings.
The error bars are significantly larger, but the central excess in modelled
neutral hydrogen column density is still present at $2\sigma$ significance.

\section{Cluster mass}\label{sec:massa}

\subsection{Gas mass}

In order to determine the cluster gas mass we assume that the gas distribution
can be described by a spherically symmetric $beta$-model profile with values of
core radius, $r_{c}$, and shape parameter, $\beta$ as determined in Paper I. With
the good quality spectrum we now dispose, it is possible to estimate the
central electronic density, $n_{0}$, as follows.

We integrate the bremsstrahlung emissivity along the line-of-sight, in the
energy bands given in Table~\ref{tbl:Sumario} and obtain $n_{0}$. Taking into
account the error bars in $r_{c}$ and $\beta$, we obtain $n_{0} = (3.7 \pm 0.6)
10^{-3}$cm$^{-3}$. Notice that the shape of the temperature profile has only a
minor effect (less than 5\%) in the determination of $n_{0}$.

The integrated gas mass profile (or growth curve) has been obtained by
integration of Eq.~(\ref{eq:betamodel}) and is shown below in
Fig.~\ref{fig:MassProfile}.

\subsection{Dynamical mass}\label{sec:mass}

With the density and temperature profiles, we can estimate the total cluster
mass. Supposing the ICM is in hydrostatic equilibrium, the total
``X-ray'' dynamical mass (i.e., the dynamical mass determined from an X-ray
observation, not to be confused with the mass of the X-ray emitting gas)
inside $r$ is given by:
\begin{equation}
    M(r) = - \frac{k T}{G\, \mu\, m_{\rm p}} \, r\, \left( \frac{\diff \ln 
    \rho(r)}{\diff \ln r} + \frac{\diff \ln T(r)}{\diff \ln r}\right) \, ,
    \label{eq:massa}
\end{equation}
where $\mu\, m_{p}$ is the mean mass per particle ($\mu = 0.59$ for a fully
ionized primordial gas) and $m_{\rm p}$ is the proton mass. We assume that
matter is distributed spherically.

Adopting the $\beta$-model to describe the intracluster gas density and the 
polytropic equation of state, the dynamical mass results in:
\begin{equation}
    M(r) = \frac{3 kT_{0} \beta \gamma r_{c}}{G \mu m_{\rm p}} \left(
    \frac{r}{r_{c}} \right)^{3} \left( 1 + 
    \left[\frac{r}{r_{c}}\right]^{2}\right)^{-1 - \frac{3}{2}(\gamma-1) 
\beta} \, .
    \label{eq:masspoly}
\end{equation}

\begin{figure}[htb]
  \centering
  \includegraphics[width=8.6cm]{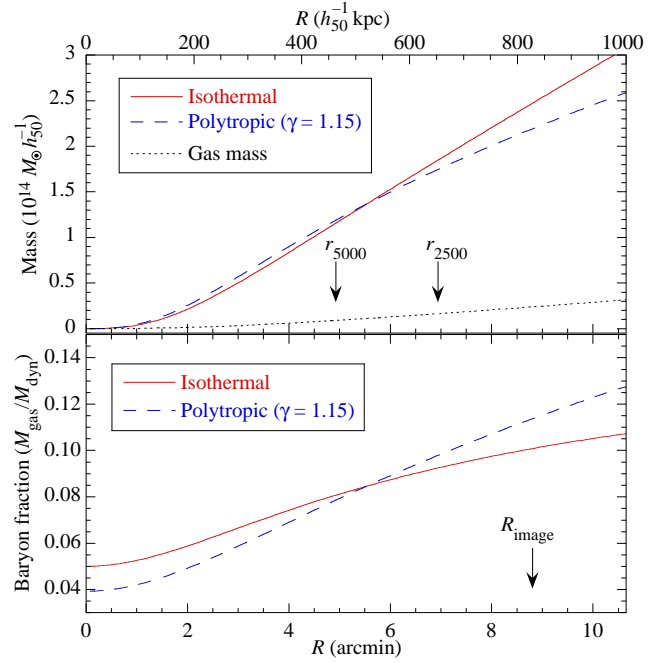}
  \caption[]{Top: X-ray dynamical mass profile using an isothermal model, with
  the mean cluster temperature (solid line), and polytropic model, with
  central temperature $T_{0} = 5.0$~keV and $\gamma = 1.15$ (dashed curve). The
  gas mass is also shown (dotted curve). Bottom: Baryon fraction given by the
  dynamical and gas mass ratio. It is shown $r_{2500}$ and $r_{5000}$, where
  the density contrast, $\rho(r)/\rho_{\rm c}$, is 2500 and 5000, respectively
  (see Appendix). $R_{\rm image}$ is the maximum radius analysed.}
  \label{fig:MassProfile}
\end{figure}

When $\gamma=1$ we have the usual expression for the isothermal sphere.
Figure~\ref{fig:MassProfile} shows the mass profile computed both with an
isothermal temperature and polytropic model. 

The dynamical masses estimated here are about 15\% to 25\% higher than those
given in Paper I where a 3.3~keV isothermal model was assumed.

We can now compute the radius corresponding to the density ratio
$\bar{\rho}(r_{\delta})/\rho_{\rm c}(z)$ (see Appendix). For $\delta = 200$,
we have the usual $r_{200}$ that is often used as an approximation to the
virial radius (e.g., \citealt{Lacey}). Assuming either an isothermal or
polytropic profile, we obtain $r_{200} = 2.3\, h_{50}^{-1}$Mpc and $r_{200} =
1.9\, h_{50}^{-1}$Mpc, respectively. Thus, $R_{\rm image}$, the maximum radius
analysed, is 0.33--$0.4\, r_{200}$, depending on the adopted temperature
profile.

From the dynamical and gas mass profiles, we can derive the baryon fraction
profile, $f(r) \equiv M_{\rm baryon}/M_{\rm total}$, assuming that the bulk of
the baryons are in the cluster ICM (i.e., $M_{\rm baryon} \approx M_{\rm
gas}$). The neglect of the galactic contribution (about $\sim$ 5 to 10\% of
the total mass, see Paper I), underestimates $f$ by just a few percent.

The bottom panel of Fig.~\ref{fig:MassProfile} gives the estimated baryon
fraction profile. For reference, we show $r_{2500}$ and $r_{5000}$, the radii
within the mean cluster density is 2500 and 5000 times the value of the
critical density of the Universe.

The gas mass fraction obtained here agrees with the typical values obtained by
\cite{WhiteFabian} (at $R = 1\,$Mpc) and by \cite{Allen} (at $R = r_{2500}$).
However, the shape of the $f(r)$ is much flatter than the profiles computed by
\citeauthor{Allen}. A possible explanation might be the adopted model for the
mass: while we assume a $\beta$-model for the gas profile, \citeauthor{Allen}
adopt the NFW profile \citep{Navarro} for the total mass.

Finally we must stress that although the new estimates of the X-ray
dynamical mass presented here are higher than those given in Paper I, they
are still a factor, $\sim 5$--10, below the virial mass estimates (VME)
obtained by the application of the Virial Theorem to the galaxies. As
discussed in Paper I, in the case of equilibrium, the dynamical mass by the
VME should not be overestimated by more than 20\%. We are thus lead to
interpret the large discrepancies between the virial theorem and the
X-ray dynamical mass estimates as evidence of a non-equilibrium state
of cluster. Indeed, in Paper I we had already remarked the incidence of
substructures in the galaxy projected distribution and of a large scale
gradient of their line-of-sight velocities, which are strong signs that the
cluster is out of equilibrium. We will discuss more in the following Section.

\section{Cluster merging}\label{sec:merge}

\begin{figure}[htb]
    \centering
    \includegraphics[width=8.6cm]{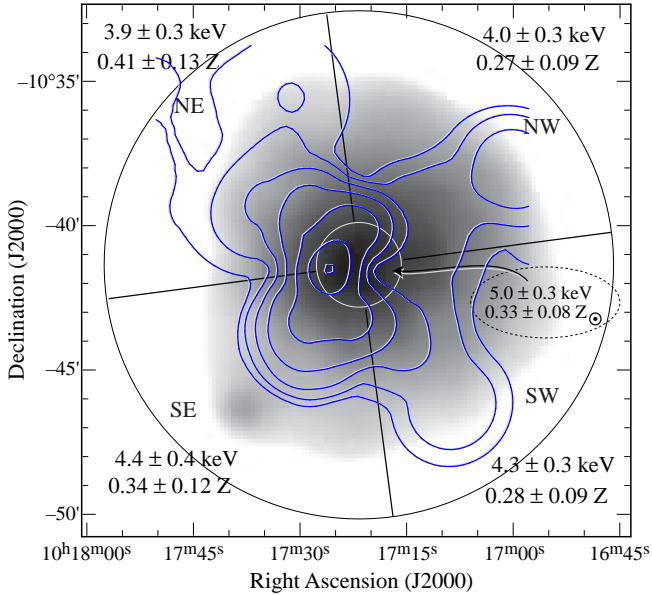}
    \caption[]{Temperature and metallicity in the regions central, NW, NE, SW,
    and SE, superposed on the combined MECS 2 and 3 adaptively smoothed image.
    The solid lines mark the cluster regions for which we carried out the
    spectroscopic analysis. The central $1'28''$ region corresponds to the
    innermost point in
    Figs.~\ref{fig:kT-R_A970Beppo}--\ref{fig:nH-R_A970Beppo}. The isocontours
    (spaced logarithmically) of the projected density map of galaxies
    kinematically bound to the cluster and with $B < 19$, are superposed to
    the X-ray image (cf. Fig~1 in Paper I).}
    \label{fig:MECS23_a970.sec}
\end{figure}

Figure \ref{fig:MECS23_a970.sec} shows the temperature and metallicity
measured in four quadrants: NW, NE, SW, and SE region. 

As it can be seen from this figure, there is a hint of a temperature gradient
in the North--South direction. Also, there is a possible metallicity gradient
in the NE--SW direction. However, within their error bars, those values are
also compatible with an isotropic distribution. In Paper I, it has been shown
that there is a large-scale gradient of the local mean velocities of the
cluster galaxies. The general direction of this velocity gradient is
approximately the same as the temperature gradient hinted in Figure
\ref{fig:MECS23_a970.sec}.

Moreover, we observe that the irregular galaxy distribution does not coincide
with the X-ray emission map. There is also a discrepancy between the dynamical
mass determined by the virial theorem and X-ray gas emission (see
Sect.~\ref{sec:massa}). These features are recognized as signs of past mergers
between sub-clusters. We may infer that, in this case, the merger is not so
recent, since the X-ray distribution already shows a fairly axial symmetric
distribution.

From another perspective, as discussed in Paper I and shown in
Fig.~\ref{fig:MECS23_a970.sec}, the galaxy distribution presents a
substructure located at a projected distance of $760\, h_{50}^{-1}$kpc ($\sim 8$
arcmin) NW of the centre, with a velocity dispersion of
$380^{+120}_{-70}$km~s$^{-1}$ inside a circular radius of $285\, h_{50}^{-1}$kpc
($\sim 3$ arcmin). This was interpreted as a group falling into the core of
Abell~970. Taking into account the mean velocities of the central region,
$17600 \pm 118$~km/s, and NW substructure, $17834 \pm 135$, (cf. Paper I),
then we can roughly assume that the substructure may be falling with a
velocity in the range 50--350~km/s. Consequently, with such a crude estimate,
we conclude that a merger will occur in the next 2--6~Gyr.

Another possibility is that this substructure has already passed near the
cluster centre, disturbing the overall cluster equilibrium, and will come back
again to the cluster central regions, before being destroyed.

Therefore, the galaxy distribution (spatial and velocity) together with the
X-ray isophote off-centre and possible temperature gradient suggest that Abell
970 is between two merger events: the first one that is responsible for the
offset between the X-ray and galaxy distributions;
the second one will occur when the NW substructure falls into the central part
of the cluster.

\section{Metal abundances}\label{sec:metal}

The pattern observed in the heavy-element abundances of the intracluster
medium provides important clues on the types of galaxies and stars producing
these elements and the mechanisms responsible for their removal from the
interstellar medium of their parent galaxies.

Using the whole 8 arcmin field (0.33--$0.4\, r_{200}$, see
Sect.~\ref{sec:mass}), we fitted a plasma model with the following metal
abundances as free parameters: Si, S, Ar, Ca, Fe, and Ni. The following
element abundances were fixed to $0.3\, Z_{\odot}$: He, C, N, O, Ne, Na, Mg,
and Al.

\begin{figure}
\centering
\includegraphics[width=8.6cm]{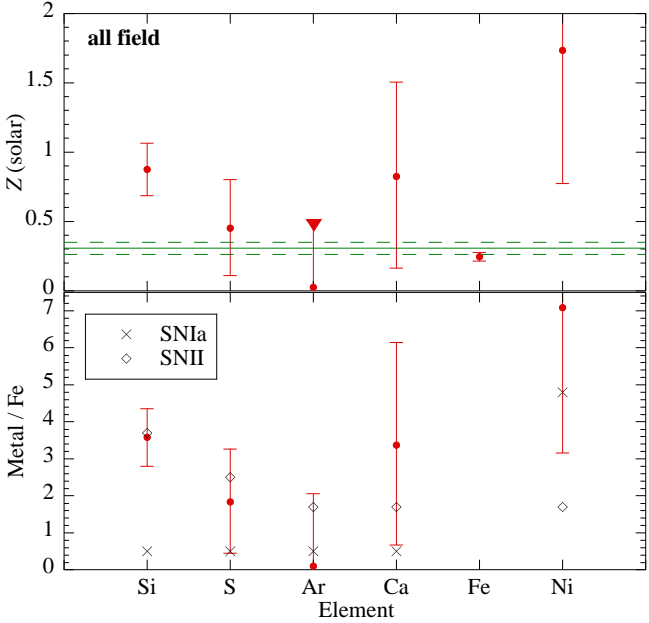}
\caption[]{Top: Individual metallicities of Fe, Ni and $\alpha$-elements
obtained with an absorbed \textsc{vmekal} model. The solid horizontal line is
the mean metallicity obtained with an absorbed \textsc{mekal} model; the
dashed lines correspond to $1\sigma$ confidence level, except for argon which
is an upper limit. Bottom: abundance ratios between metal and iron. Also
shown are the predicted approximate yields of Type II SNe (diamonds) and Type Ia
SNe (crosses), based on \cite{Dupke}.}
\label{fig:metal_indivA970SAX}
\end{figure}

The result is summarized in Figure \ref{fig:metal_indivA970SAX}, which shows
the over-abundance of $\alpha$-elements relative to iron. Both the silicon and
sulphur abundances are consistent with the typical values found recently by
\citet{Finoguenov2} and are slightly above the value determined by
\citet{Fukazawa} for a cluster of $\sim 4.5$~keV (which is $\sim 0.6$--$0.8
Z_{\odot}$, cf. Fig.~2 of their paper). These authors have shown that the Si
abundance correlates with the gas temperature, probably due to a decreasing
contribution from Type II supernovae (SNe) in poorer clusters, where the
metals produced by this type of SN could escape from the potential well.

The over-density of $\alpha$-elements displayed in
Fig.~\ref{fig:metal_indivA970SAX} is characteristic of the Type II SNe yield.
This suggests that the intracluster gas enrichment was caused by ejected gas
that was synthesized in massive stars, the progenitors of Type II SNe. Present
day elliptical galaxies show a very low rate of Type II SNe, implying that the
enrichment of the intracluster gas probably occurred early in the history of
the cluster.

However, we cannot discard other possible enrichment mechanisms. In fact, as
shown in Figure~\ref{fig:metal_indivA970SAX}, the abundance ratio Ni/Fe is
higher than expected for Type II SNe, but it is in agreement with the expected
Type Ia SNe yield. This finding supports models where part of the metals in
the ICM were produced by Type Ia SNe and which were either blown from the
galaxies by SNe explosions or swept by ram-pressure (e.g., \cite{Dupke}).
However, the Ni abundance is uncertain. Moreover, the abundance ratio Si/Fe
(with its smaller error bars) is incompatible with enrichment by Type Ia SNe.

\section{Conclusions}

We have presented here a new X-ray observation of Abell 970. Below we
summarize our main results.

\begin{itemize}
    
 \item Although Abell 970 presents rather symmetric X-ray isophotes (except
 for an excess emission that is probably due to an active galactic nucleus,
 triggered by a galaxy merger), the peak of the X-ray distribution does not
 coincide with the centre of the galaxy distribution.
 
 \item The temperature profile presents a strong radial gradient at $\sim 6$
 arcmin, which is barely compatible with adiabatic stable equilibrium. The
 metal abundances are approximately constant up to 8~arcmin. There are
 indications of non radial temperature and metallicity gradients (but with
 confidence levels less than $1 \sigma$), notably in the NW--SE direction,
 the same direction where a gradient in velocity is observed (cf. Paper I).
 
 \item With the two findings above, we arrive at two possible scenarios for the
 recent history of Abell~970: either it has suffered a recent merger with a
 subcluster, or the NW substructure (detected only from galaxy counts) has
 recently passed through the centre of Abell~970. In both cases, the dynamical
 stability was disrupted, but the cluster is now approaching virial
 equilibrium again. The observed galaxy distribution suggests that this
 equilibrium will be broken in a few Gigayears when the nearby NW substructure
 will fall (again?) into the core of Abell~970.
 
 \item We have derived an abundance ratio between $\alpha$-elements/iron
 greater than 1 and the iron metallicity is lower than the mean metallicity.
 The over abundance of $\alpha$-elements may be an indication of an early
 enrichment of the ICM by Type II supernovae in elliptical galaxies. Part or
 the metals in the ICM could be produced by Type Ia SNe, but the evidence from
 the Ni/Fe ratio is not strong.

 \item We have found a central X-ray absorption excess compared to the
 absorption expected from the mean galactic value of $N_{\rm H}$, although the
 statistical error bars are large. The central $N_{\rm H}$ excess could be
 associated to the central galaxies and/or to cooled intra-cluster material
 (either in molecular form and/or dust grains). The latter may be preferable
 since the core of the cluster is dominated by early-type galaxies. However,
 this result probably depends on the spectral modelling \citep[see,
 e.g.,][]{McCarthy02}.

\end{itemize}

\begin{acknowledgements}
We would like to thank Gary Mamon and an anonymous referee for a critical
reading of this paper. We acknowledge support from the COFECUB
(\textit{Comit\'e Fran\c{c}ais pour l'Evaluation de la Coop\'eration avec le
Br\'esil}), CNPq (\textit{Conselho Nacional de Desenvolvimento Cient\'{\i}fico
e Tecnol\'ogico}), \textsc{fapesp} (\textit{Funda\c{c}\~ao de Amparo \`a
Pesquisa do Estado de S\~ao Paulo}), \textsc{pronex}, and the PICS
France--Brazil cooperation. DP acknowledges IAG/USP for its hospitality. This
research has made use of SAXDAS linearized and cleaned event files (Rev.1.1)
produced at the BeppoSAX Science Data Center.
\end{acknowledgements}

\appendix
\section{Derivation of $r_{\delta}$}\label{sec:apendice}

In Paper I, we give a formula for $r_{\delta}$, which is defined by:
\begin{equation}
    \delta \equiv \frac{\bar{\rho}(r_{\delta})}{\rho_{\rm c}(z)} \, ,
    \label{eq:delta}
\end{equation}
where $\bar{\rho}(r_{\delta}) \equiv M_{\rm dyn}(r_{\delta})/(4 \pi
r_{\delta}^{3}/3)$ is the mean density of the cluster inside $r_{\delta}$ and 
$\rho_{\rm c}(z)$ is the critical density of the Universe at the redshift of 
the cluster.

We take the dynamical mass corresponding to a $\beta$-model ICM and polytropic 
temperature profile, Eq.~(\ref{eq:masspoly}), and the critical density given 
by:
\begin{equation}
     \rho_{\rm c}(z) = \displaystyle{\frac{3\, H^{2}(z)}{8 \pi\, G}} \, ; 
     \quad H(z) = H_{0} f(z, \Omega_{M}, \Omega_{\Lambda}) \, ,
    \label{eq:rhocrit}
\end{equation}
with
\begin{equation}
    \begin{array}{l}
      f^{2}(z, \Omega_{M}, \Omega_{\Lambda}) = \\
      \qquad \Omega_{\Lambda} + \Omega_{M}(1+z)^{3} - 
    (\Omega_{\Lambda} + \Omega_{M} - 1)(1 + z)^{2} \, .
    \end{array}
    \label{eq:fz}
\end{equation}
Combining these equations and solving for $r_{\delta}$ we obtain:
\begin{equation}
 \begin{array}{l}
 \displaystyle{\left(\frac{r_{\delta}}{r_{c}}\right)^{2} + 1 =} \\
 \qquad\quad \displaystyle{\left[\frac{2.3\times 
     10^{8}}{\delta \, h_{50}^{2} f^{2}(z, \Omega_{M}, \Omega_{\Lambda})}
     \frac{\beta \, \gamma \, T_{0}}{\mu r_{c}^{2}}
     \right]^{1/[1+3\beta(\gamma-1)/2]}} \, ,
 \end{array}
 \label{eq:rdelta}
\end{equation}
where the cluster central temperature, $T_{0}$, is given in keV and the core 
radius, $r_{c}$, in kpc. For $\delta \la 500$, $r_{\delta} \gg r_{c}$, and 
Eq.~(\ref{eq:rdelta}) is simplified:
\begin{equation}
    \frac{r_{\delta}}{r_{c}} \simeq \left[\frac{2.3\times 
    10^{8}}{\delta \, h_{50}^{2} f^{2}(z, \Omega_{M}, \Omega_{\Lambda})}
    \frac{\beta \, \gamma \, T_{0}}{\mu r_{c}^{2}}
    \right]^{1/[2+3\beta(\gamma-1)]}  .
    \label{eq:simplif}
\end{equation}
When $\gamma=1$, the isothermal case is obtained and $T_{0}$ is the mean 
temperature.

For $z \ll 0.1$ we can use the approximation:
\begin{equation}
    f^{2}(z, \Omega_{M}, \Omega_{\Lambda}) \simeq
    1 + (2 - 2 \Omega_{\Lambda} + \Omega_{M}) \, z \, .
    \label{eq:fzaprox}
\end{equation}

\end{document}